\documentclass[aps,prb,twocolumn,floatfix,footinbib,showpacs]{revtex4}
\usepackage{graphicx}
\usepackage{amsfonts,amsmath,amssymb}
\usepackage{amsthm}
\usepackage{dsfont,bm}
\usepackage{color}
\usepackage{soul} 
\usepackage[colorlinks=true,linkcolor=blue,pagecolor=blue,filecolor=blue,menucolor=blue,urlcolor=blue,citecolor=blue,anchorcolor=blue]{hyperref}%
\usepackage{sidecap}

\begin{document}

\title{Tunable Anomalous Andreev Reflection and Triplet Pairings in Spin Orbit Coupled Graphene}

\author{Razieh Beiranvand }
\affiliation{Department of Physics, K.N. Toosi University of Technology, Tehran 15875-4416, Iran}
\author{Hossein Hamzehpour}
\affiliation{Department of Physics, K.N. Toosi University of Technology, Tehran 15875-4416, Iran}
\author{Mohammad Alidoust}
\affiliation{Department of Physics, K.N. Toosi University of Technology, Tehran 15875-4416, Iran}


\date{\today}

\begin{abstract}
We theoretically study scattering process and superconducting triplet correlations in a graphene junction comprised of ferromagnet-RSO-superconductor in which RSO stands for a region with Rashba spin orbit interaction. Our results reveal spin-polarized \textit{subgap} transport through the system due to an anomalous equal-spin Andreev reflection in addition to conventional back scatterings. We calculate equal- and opposite-spin pair correlations near the F-RSO interface and demonstrate direct link of the anomalous Andreev reflection and \textit{equal-spin} pairings arised due to the proximity effect in the presence of RSO interaction. Moreover, we show that the amplitude of anomalous Andreev reflection, and thus the triplet pairings, are experimentally controllable when incorporating the influences of both tunable strain and Fermi level in the nonsuperconducting region. Our findings can be confirmed by a conductance spectroscopy experiment and provide better insights into the \textit{proximity-induced} RSO coupling in graphene layers reported in recent experiments \onlinecite{avsar2014nat,ex1,ex2,ex3}.
\end{abstract}
\maketitle
\section{introduction}
Ferromagnetism and $s$ wave superconductivity are two phases of matter with incompatible order parameters. The interplay of superconductivity and ferromagnetism in a junction platform results in intriguing and peculiar phenomena \cite{Buzdin2005RMP,eschrig_2015,bergeret_rmp}. For instance, in a standard uniform superconductor-ferromagnet-superconductor (S-F-S) junction, supercurrent can go under multiple reversals when varying temperature, exchange field, and junction thickness \cite{Ryazanov2001PRL}. These sign reversals are due to the apparence of triplet opposite-spin pairings (OSPs) and thus the oscillation of Cooper pairs' amplitude in the time-reversal broken region i.e. F \cite{Buzdin2005RMP,kupriyanov,bergeret_rmp,ma_jap}. If magnetization in the F region follows a nonuniform pattern a new type of superconducting correlations arises: Triplet equal-spin pairings (ESPs). The equal-spin pair correlations are long range and can extensively propagate in materials with uniform magnetization and strong scattering resources \cite{Buzdin2005RMP,eschrig_2015,bergeret_rmp,ma_jap}. This long range nature of spin-polarized superconducting correlations has turned the ESPs to a highly attractive perspective in nanoscale spintronics \cite{eschrig_2015,rob,bob,moor,sat,khay,sac,halfm}. Another source to generate the ESPs is a combination of spin orbit interactions and uniform Zeeman field proximitized to a superconducting electrode \cite{Hogl2015PRL,bergprb,ma_jpc,ma_njp}. One of the main advantages of making use of spin orbit interactions to induce the ESPs is an all electrical control over the ESPs \cite{bobkova,bergprb,Hogl2015PRL,Konschelle,grigor,ma_jpc,ma_njp,arijit}.

Graphene is a single layer of carbon
atoms, arranged in hexagonal lattices, with a linear dispersion 
at low energies and tunable Fermi level that can be simply manipulated by a gate voltage \cite{cast1,beenakkerrmp}. These exceptional characteristics in addition to a long spin relaxation time of moving charged carriers, compared to their counterparts in a standard conductor, has turned graphene to a promising material for spintronics devices \cite{cast1,beenakkerrmp}. Superconductivity, ferromagnetism, and
spin orbit interactions can be induced into graphene layers by means of the proximity effect \cite{Heersche2007Nature,Tombros2007Nature,Chakraborty2007PRB,Dedkov2008PRL,Varykhalov2012PRL,vahid,rameshti,ex1,ex2,ex3,avsar2014nat}. It was experimentally demonstrated that a graphene monolayer can support strong Rashba spin orbit interaction of order of $\sim$ 17 meV by proximity to a semiconducting tungsten disulphide substrate \cite{avsar2014nat}. Recent developments have achieved large proximity-induced ferromagnetism and spin orbit interactions in CVD grown graphene single layers coupled to an atomically flat yttrium iron garnet \cite{ex1,ex2,ex3}. To demonstrate the existence of proximity-induced RSO interaction in the graphene layer, a DC voltage along the graphene layer is measured by spin to charge current conversion which is interpreted as the inverse Rashba-Edelstein effect \cite{ex1,ex2,ex3}. The Edelstein effect was first discussed in connection with the spin polarization of conduction electrons in the presence of an electric current \cite{ed}. Furthermore, graphene has the capability of sustaining strain and deformations without rupture \cite{Peres2010RMP,Pereire2009PRB}. The application of strain to graphene layers can result in important and interesting phenomena \cite{Jiang_strain,Assili_strain1,Verbiest_strain1,Benjamin_strain1,Manes_strain,Gruji_strain,Wang_strain1,strain,ramin,sattari}. For example, the interplay of massive electrons with spin orbit coupling in the presence of strain in a graphene layer yields controllable spatially separated spin-valley filtering \cite{Gruji_strain,Jiang_strain}. Therefore, this property can be employed as a means to control the spin-transport graphene-based spintronics devices \cite{ramin}.
 
In this paper, we incorporate RSO interaction, superconductivity, ferromagnetism, and two different types of strain in a set up of graphene-based F-RSO-S contact (depicted in Fig. \ref{fig1}) and propose an experimentally feasible device to generate `controllable' odd-frequency superconducting triplet correlations. Our results reveal a finite anomalous equal-spin Andreev reflection due to the RSO interaction. We demonstrate that this anomalous reflection results in nonvanishing superconducting ESPs in the ferromagnetic region near the F-RSO interface. We also vary the Fermi level (that can be experimentally achieved using a gate voltage) and the strength of an applied strain and show that by simply tuning these two physical quantities one can suppress the amplitude of opposite-spin superconducting correlations while simultaneously increase the equal-spin pairings at the F-RSO interface. Furthermore, we study the charge and spin conductances of the junction and show that the anomalous equal-spin Andreev reflection yields a finite tunable spin-polarized subgap conductance which is experimentally measurable. Such a spectroscopy experiment can be an alternative to those of Refs. \onlinecite{avsar2014nat,ex1,ex2,ex3} for demonstrating the existence of proximity-induced RSO in graphene layers and determining its characteristics. We complement our findings by investigating the system's band structure and discussing its aspects.

The paper is organized as follows. We first summarize the theoretical framework used in Sec. \ref{theor}. The results are discussed in Sec. \ref{disc} in three subsections: in Subsec. \ref{bndstr} we study the band structure of system and discuss various reflection and transmission probabilities, in Subsec. \ref{esp} the explored anomalous Andreev reflection is linked to the ESPs, and we study the spin and charge conductances in Subsec. \ref{cond}. More details of analytics and calculations are discussed in Appendix. We finally summarize concluding remarks in Sec. \ref{concld}.

\section{formalism and theory}\label{theor}

\begin{figure}[t!]
   \includegraphics[width=7.5cm, height=4.5cm]{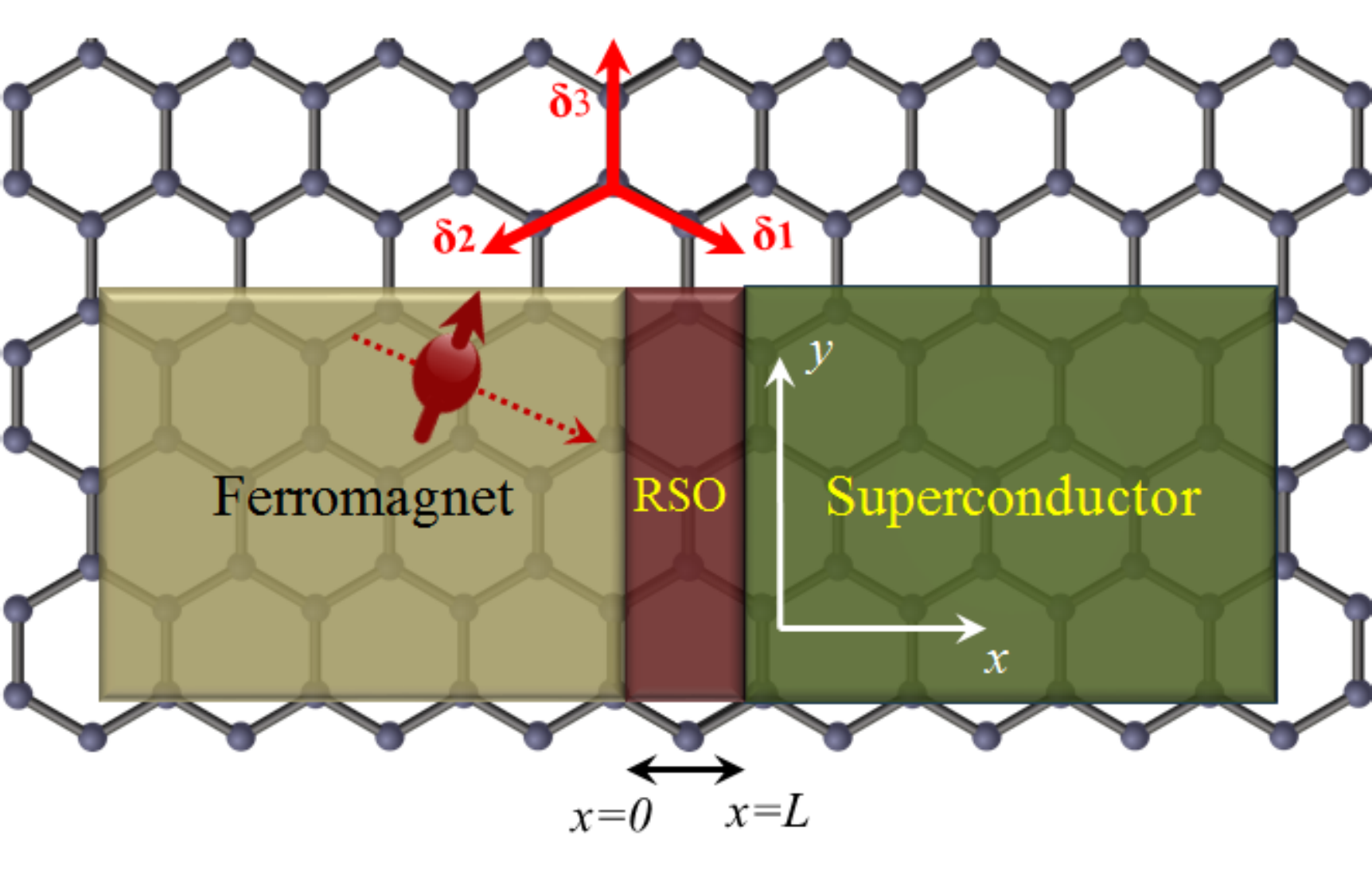}
   \caption{Schematic of the graphene-based F-RSO-S junction. The junction resides in the $xy$ plane and the RSO region has a thickness of $L$ that extends from $x=0$ to $L$. The uniform ferromagnetic and superconducting regions are semi-infinite and constitute interfaces with the RSO region at $x=0$ and $x=L$. We denote the displacement unit vectors of graphene unit cells by ${\bm \delta}_{1,2,3}$. A possible trajectory of spin-up particles incident at the F-RSO interface within the F region is shown.}
   \label{fig1}
    \end{figure}

Because of the specific band shape of monolayer graphene in low energies, carriers in
such a system can be treated as massless Dirac fermions
\cite{cast1}. Thus, to describe the low-energy excitations of
the structure shown in Fig. \ref{fig1}, one can incorporate the Dirac Hamiltonian with
the Bogoilubov-de Gennes equation to reach at a Dirac-Bogoliubov-de Gennes
(DBdG) equation in the presence of RSO coupling and exchange field as follow \cite{Beenakker}:
 \begin{equation}
       \left(\begin{array}{cc}
       \mathcal{H}_\text{D}+\mathcal{H}_i-\mu^i & \Delta\\
       \Delta ^{\star} & \mu^i-{\cal T}[\mathcal{H}_\text{D}-\mathcal{H}_i]{\cal T}^{-1}\\
       \end{array}\right)
       \left(\begin{array}{c}
        u \\
        v \\
       \end{array}\right)=\varepsilon
       \left(\begin{array}{c}
       u \\
       v \\
       \end{array}\right),
       \label{Eq.DBdG}
   \end{equation}
  where $\varepsilon$ is the quasiparticles' energy and $u $ and $v$ refer to the electron and hole parts of
  spinors, respectively. $\mathcal{H}_\text{D}=s_0\otimes\left(\sigma_x v_x^i k_x+\sigma_y v_y^i k_y\right)$ is a two dimensional massless Dirac Hamiltonian which governs low energy excitations in one valley of graphene and ${\cal T}$ is the time reversal operator \cite{beenakkerrmp}. 
$k_x$ and $k_y$ are components of wave vector in the $x$ and $y$ directions, respectively. $\sigma_i$ and $s_i$ are Pauli matrices, acting on the pseuedospin and real spin spaces of graphene ($\sigma_0$ and $s_0$ are $2\times 2$ unit matrices) and natural units are used: $\hbar=1$. The index $i$ labels F, RSO, or S regions as seen in Fig. \ref{fig1}:
    \begin{equation}
     \mathcal{H}_i(x)=
      \begin{cases} 
           \mathcal{H}_\text{F}=\left( s_z\otimes\sigma_0\right) h & x\leq 0 \\
           \mathcal{H}_\text{RSO}=\lambda \left( s_y\otimes\sigma_x-s_x\otimes\sigma_y\right) & 0\leq x\leq L \\
           \mathcal{H}_\text{S}=-U_0 s_0\otimes\sigma_0 & x\geq L 
        \end{cases}
        \label{Eq.Hnet},
     \end{equation} 
where $L$ indicates the thickness of RSO
region. Here $\lambda$ is the energy scale of spin orbit coupling and $U_0$ is the electrostatic potential in the superconducting region. Because of valley
degeneracy in a single layer of graphene, one can simply multiply
final results by a factor of two. 
In a graphene layer under strain $v_x^i\neq v_y^i$ which implies anisotropic Fermi
velocity in F, RSO, and S regions.
The Fermi energy in each region is shown by $\mu^i$. In the Hamiltonian of F segment, Eq.(\ref{Eq.Hnet}), $h$ represents the
exchange field which is added to the Dirac Hamiltonian via the Stoner
approach. For simplicity in our calculations, we assume that $h$ is
oriented along the $z$ direction
without loss of generality \cite{halt1}. This choice turns the
exchange field to a good quantum number that allows for explicitly
considering spin-up and -down quasiparticles in the F region and helps
having insightful analyses of spin-dependent phenomena in the system. The superconducting gap $\Delta$ is a
matrix in the particle-hole space (nonzero in $L\leq x$) and is given by:
\begin{figure*}
\includegraphics[width=18.cm,height=6.7cm]{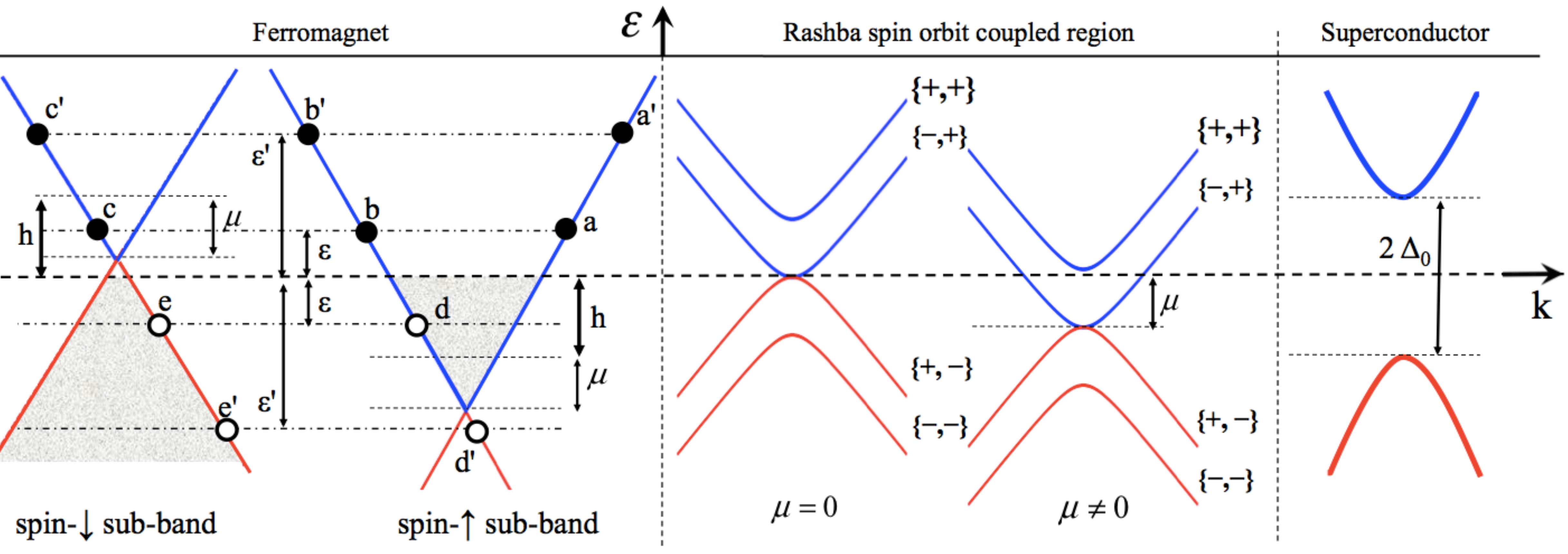}
\caption{(Color online) Band structure of each region. The left panel
  shows the band structure of F region, illustrating different reflection
  possibilities including retro and specular reflections discussed in the text. The circles stands for the holes while the solid circles stand for the electrons. The middle panel is the band structure of RSO
  region at the charge neutrality point and away from it i.e. $\mu=0$ and $\mu\neq 0$, respectively. The $\pm$ signs
  refer to $\zeta$ and $\eta$ in Eq. (\ref{EigenHSO}) due to the band splitting effect of RSO interaction. The right panel is
  the band structure of a dopped superconductor.}
\label{bands}
\end{figure*}
  \begin{equation}
  \Delta=\Theta(x-L)\left(\begin{array}{cccc}
  \Delta_0 e^{i \phi} & 0 & 0 & 0 \\
  0 & \Delta_0 e^{i \phi} & 0 & 0 \\
  0 & 0 & \Delta_0 e^{i \phi}& 0 \\
  0 & 0 & 0 & \Delta_0 e^{i \phi}\\ 
  \end{array}\right),
  \label{Eq.Delta}
  \end{equation}
in which $\Delta_0$ is the superconducting gap at zero temperature,
$\phi$ is the macroscopic phase of superconductor, and $\Theta$
denotes a Heaviside step function. This step function assumption made is valid as far as the Fermi wavelength in the
S region is much smaller than F and RSO regions
i.e. $\lambda_F^\text{S} \ll \lambda_F^\text{F},\lambda_F^\text{RSO}$. Otherwise, a
self-consistent approach is favorable to accurately determine the spatial
profile of the pair potential \cite{halt_2011,halt1}. We also assume that the F-RSO junction can be described by a step change from the ferromagnetic region to RSO. Although we initially do not consider a smooth change at this junction (that can happen in realistic systems due to the proximity effect), each region eventually gains its own neighbour properties near the boundary by matching their wavefunctions at this location \cite{beenakkerrmp}. We note that such modifications, including weak nonmagnetic impurities and moderately rough interfaces, can only alter the amplitude of scattering probabilities and not the conclusions of our work.

To describe a strained graphene layer, we follow
Ref. \onlinecite{Pereire2009PRB}. Expanding the tight-binding model band
structure with arbitrary hopping energies around the Dirac point,
one finds: 
\begin{equation}
\epsilon=\pm \left|\sum_{i=1}^{3} t_i e^{-i {\bm k}.{\bm \delta}}\right|,
\label{Eq.23}
\end{equation}
where ${\bm \delta}$ is the lattice vector as depicted in Fig. \ref{fig1}. The position of one of the Dirac points $\mathbf{K_D}$ is
$\left( \cos^{-1}(-1/2\eta)/\sqrt{3 a_x}, 0\right)$. Here, we assume
$t_{1,2}=t_\vdash$, $t_3=t$ and $\eta=t_\vdash/t$ in our
calculations (see Ref. \onlinecite{Alidoust2011}). These assumptions constitute asymmetric velocities to the Dirac fermions in different directions.

To satisfy the mean field approximation in the S region (which is experimentally relevant) namely, the Fermi wavevector in the superconductor should be much larger than its F and RSO counterparts $k_F^{\text{S}} \gg k_F^{\text{F}}, k_F^{\text{RSO}} $ \cite{beenakkerrmp}, 
we consider a heavily dopped superconductor, that is achieved by $U_0\gg \varepsilon, \Delta_0$ \cite{Beenakker}.
The resulting wavefunctions are $1\times 8$ spinors (see Appendix). We match these wavefunctions at the boundaries i.e. at $x=0$ and $x=L$ and calculate various probabilities of the electron-hole scatterings. We normalize energies by the superconducting gap at zero temperature $\Delta_0$ and lengths by the superconducting coherent length $\xi_S=\hbar v_F/\Delta_0$. The gap of superconductor depends on the temperature ($T$) and we consider $T=0.01 T_c$ throughout our calculations in which $T_c$ is the critical temperature of superconductor. We also set the length of RSO region fixed at $L=0.2\xi_S$.


\section{results and discussions}\label{disc}
In this section we present main results of the paper. 
\begin{figure*}
\includegraphics[width=18.cm,height=7.5cm]{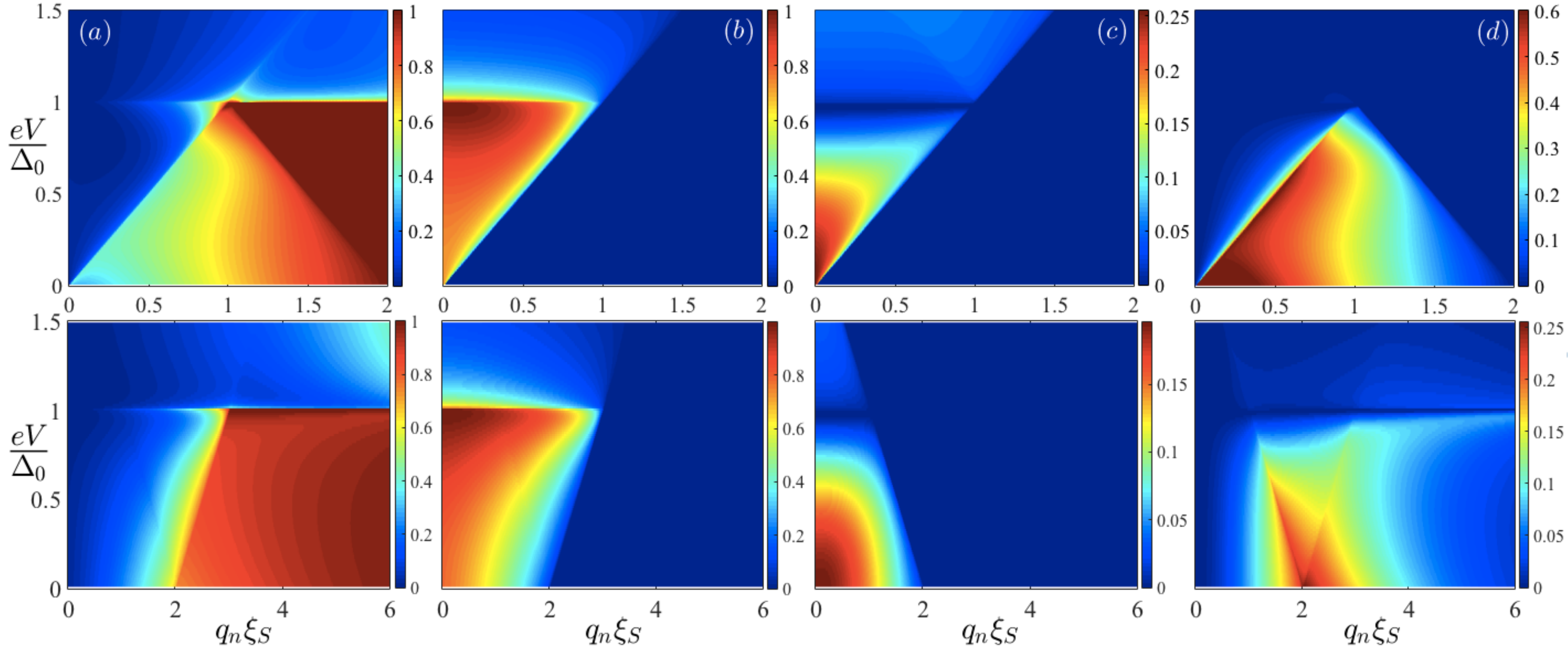}
\caption{(Color online)  Back scattering probabilities of an incident spin-up particle at $x=0$ as a function of applied voltage $eV$ across the junction and the transverse momentum $q_n\xi_S$. ($a$) conventional normal
  reflection $|r_{N}^\uparrow|^2$ , ($b$) conventional Andreev
  reflection $|r_{A}^\downarrow|^2$, ($c$) spin-flipped normal
  reflection $|r_{N}^\downarrow|^2$, and ($d$) anomalous Andreev
  reflection $|r_{A}^\uparrow|^2$ vs the transverse component of
  wavevector $q_n$. The
  vertical axis is the applied voltage across the junction. Top row: $\mu=h=\Delta_0$, and $\lambda=1.5 \Delta_0$. Bottom row: $\mu=0.5\Delta_0$, $h=7.0\Delta_0$, and $\lambda=1.5 \Delta_0$.}
\label{DominantAAKR}
\end{figure*}

 \subsection{II. Band structure and reflection probabilities}\label{bndstr}

The electronic band structure of each region can provide helpful
insights into the properties of system, particularly the various
backscatterings and superconducting correlations. To obtain
dispersion relations and corresponding spinors in each region, we diagonalize the DBdG Hamiltonian Eq. (\ref{Eq.DBdG}). The system band structure at low energies within F, RSO, and S regions can be expressed by:
  \begin{equation}
  \varepsilon=\pm \mu^\text{\tiny F} \pm \Big[ (v_x^\text{\tiny F}k^\text{\tiny F}_x)^2+(v_y^\text{\tiny F}k_y)^2\Big]^{1/2}\pm h,
  \label{Eq.04}
  \end{equation} 
  \begin{equation}
  \varepsilon=\pm\mu^\text{\tiny RSO}+\zeta \Big[(v_x^\text{\tiny RSO}k_x^\text{\tiny RSO})^2+(v_y^\text{\tiny RSO}k_y)^2+\lambda^2\Big]^{1/2}+\eta \lambda,~~~
  \label{EigenHSO}
  \end{equation} 
where $\eta,\zeta=\mp\pm 1$ and 
  \begin{equation}
  \varepsilon=\Big[|\Delta_0|^2+\Big(\mu^\text{\tiny S}+U_0\pm \sqrt{(v_x^\text{\tiny S} k_x^\text{\tiny S})^2+(v_y^\text{\tiny S} k_y)^2}\Big)^2\Big]^{1/2}.
  \label{Eq.17}
  \end{equation}
Figure \ref{bands}
illustrates the excitation spectrums in the F, RSO, and S regions. The superconductor is assumed highly dopped so that the low
energy excitation spectrum is a parabola. In our calculations of the
reflection and transmission probabilities we consider a scenario where an electron with
spin-up hits the F-RSO interface. Due to the tunable Fermi level in a graphene layer, one can consider three regimes: 
($i$) undopped regime with $\mu=0$, ($ii$) low dopped limit with
$\mu \approx \Delta_0$ and ($iii$) heavily dopped limit with
$\mu\gg\Delta_0$. In the RSO region, the band structure
has two subbands considering $\eta$ and $\zeta$ ($\{\pm ,\mp\}$) signs. Due to the subbands,
the RSO region serves as a spin mixer in the transport
mechanism. Therefore, when an electron with spin-up hits the F-RSO
interface, there is a finite probability for a hole reflection with
spin-up. The influence of $\mu$ and $h$ are shown in the F region. The solid circles show the holes in the bands and solid circles stand for the electrons. For a subgap electron with spin-up and $\varepsilon<\mu+h$, corresponding backscattering possibilities are labeled by $a$-$e$. The backscattering of $a$$\rightarrow$$b$ is the normal electron reflection, $a$$\rightarrow$$c$ is the normal spin-flipped, $a$$\rightarrow$$d$ is the anomalous Andreev reflection where the backscattered hole lies in the conduction band, and $a$$\rightarrow$$d$ is the conventional Andreev reflection where the backscattered hole passes through the valance band. We also show a case where the energy of incident electron takes a value of $\varepsilon>\mu+h$ by $a'$-$b'$ labels. In this case, the anomalous Andreev reflected hole passes through the valance band.   

In an undopped graphene layer, the chemical potential is vanishingly
small $\mu\approx 0$. Depending on the energy of incident particle, a
new type of Andreev reflection can take place which is called specular
Andreev reflection \cite{Beenakker}. The specular Andreev
reflection occurs when an electron in the conduction band is converted
into a hole in the valence band upon the scattering process. It shows
the possibility of an unusual electron-hole conversion in the
reflection of relativistic electrons in graphene junctions \cite{Beenakker}. In the
retro Andreev reflection however electron and hole both lie in the
conduction band as shown in Fig. \ref{bands}. To gain more insights into various reflections that shall be discussed below, we have presented details of calculations and reflections in Appendix.
In the low dopped regime ($\mu\approx\Delta_0$), by tuning the system
parameters, novel effects can occur:
($1$) $\mu=h\approx\Delta_0$: 
In this limit, if the RSO parameter is zero ($\lambda=0$), the Rashba
region acts similarly to a normal region with a finite width where the
particles can experience resonances upon multiple reflections from
boundaries. Hence, there are only two reflection probabilities
present: $a$) conventional normal reflection ($r_{N}^\uparrow$) and
$b$) either retro Andreev reflection ($\varepsilon \leq \mu+h$) or specular
Andreev reflection ($\varepsilon \geq \mu-h$), depending on the
quasiparticle's energy discussed above. The conventional normal reflection dominates
at energies below the superconducting gap ($\varepsilon \leq \Delta_0$)
\cite{Beenakker}. In this regime, because $\mu$
and exchange field $h$ are equal, the population of spin-down
electrons is minority and thus $r_{N}^\downarrow\sim 0$ and
$r_{A}^\downarrow$ has a finite probability.
($2$) At nonzero values of $\lambda$, a spin-up particle arriving from the
F region, can undergo a spin mixing process in the RSO segment. In this limit,
in addition to the normal reflection $r_{N}^\uparrow$, the
probabilities of anomalous Andreev reflection ($r_{A}^\downarrow$) and
unconventional normal reflection ($r_{N}^\downarrow$) have finite amplitudes (see Appendix).
Using this choice of parameters, the conventional Andreev
reflected hole is placed in the valence band while the anomalous one
passes through the conduction band. Moreover, a hole in the conduction band
belongs to the retro-reflection process whereas a hole in the valence band
belongs to the specular reflection mechanism \cite{Beenakker}. Thus, the reflected
holes can follow two different processes inside the F region depending on their spin orientation (see left panel of Fig.\ref{bands}). 
In a heavily dopped graphene $\mu\gg\Delta_0$, the Andreev
reflection is of retro-type. The propagation of
carriers are limited by critical angles that can describe their incident
or reflection angles:
\begin{equation}\label{eqcang}
\begin{cases}
\alpha_{e\downarrow}^c=\arcsin(\dfrac{\varepsilon+\mu-h}{\varepsilon+\mu+h})\\
\alpha_{h\downarrow}^c=\arcsin(\dfrac{\varepsilon-\mu+h}{\varepsilon+\mu+h})\\
\alpha_{h\uparrow}^c=\arcsin(\dfrac{\varepsilon-\mu-h}{\varepsilon+\mu+h})
\end{cases}.
\end{equation}
As seen, in a regime where $\mu=h\gg \Delta_0$, $\varepsilon$, the
critical angles $\alpha_{e,\downarrow}^c \approx
\alpha_{h,\downarrow}^c\approx 0$ vanish. Therefore, their corresponding
probabilities do not contribute to the quantum transport. Our investigations demonstrate that in a
certain regime of parameter space where $\lambda=\mu=h$, the anomalous
Andreev reflection highly dominates in the $eV-q_n$ space. This regime results in zero conventional Andreev reflection while at $\lambda\geq h$ the probability of anomalous Andreev reflection at the edge of superconducting gap becomes unity. This effect suggests a spin-polarized Andreev-Klein reflection \cite{zar1}. Note
that the existence of phenomena described above are dependent on the
presence of $\lambda$.  

        \begin{figure}[t]
        \includegraphics[width=8.cm, height=6.cm]{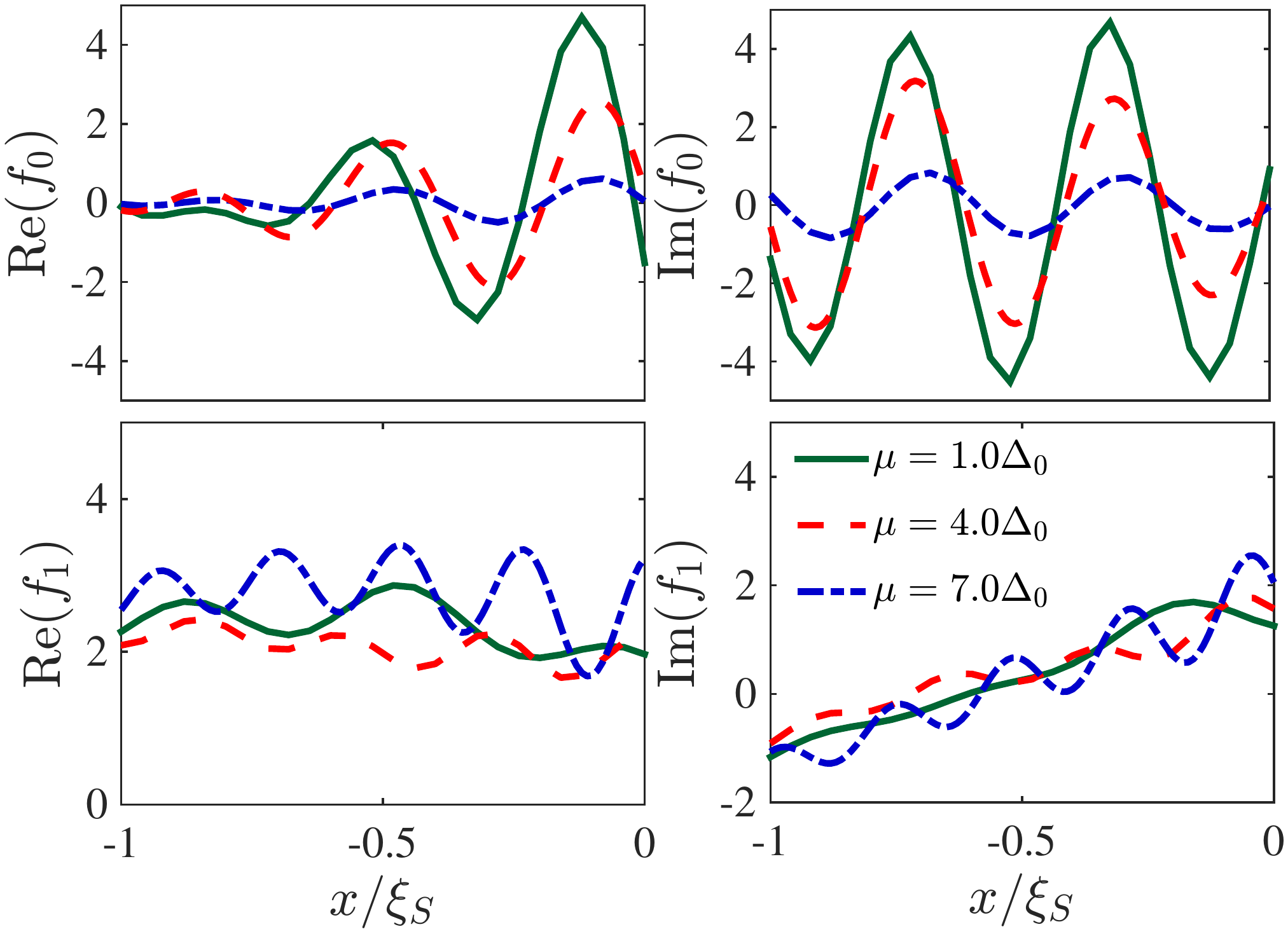}
        \caption{(Color online) Real and imaginary parts of OSP ($f_0$) and ESP ($f_1$) as a function of position in the F region $x \leq 0$. The Fermi energy is set at $\mu=1.0\Delta_0, 4.0\Delta_0$, and $7.0\Delta_0$ and the voltage difference is assumed constant at $eV=0.5\Delta_0$. The other parameters are the same as those of Fig. \ref{DominantAAKR} bottom row. }
        \label{fig3}
        \end{figure} 
Figure \ref{DominantAAKR} exhibits back scattering probabilities of an incident particle with spin-up at the F-RSO interface. Throughout our calculations, we consider a fairly narrow region that allows more clear analysis of the ESPs and anomalous Andreev reflection. In the top row panels we set the chemical potential at the superconducting gap $\mu=\Delta_0$ and plot probabilities as a function of an applied voltage across the junction $eV/\Delta_0$ and the transverse particle's momentum $q_n\xi_S$ which is a conserved quantity during the scattering process throughout the system. Here we also consider $v_x^i=v_y^i$ corresponding to zero strain exertion into the system. As seen, in the presence of RSO region, an Andreev reflected hole with the same spin direction as the incident particle $|r_{A}^\uparrow|^2$ can occur with a finite probability. Furthermore, by calibrating $\mu$, spin orbit coupling strength $\lambda$, and strain one can suppress the conventional Andreev reflection and simultaneously generate anomalous Andreev reflected holes with a finite probability. We note that $|r_{A}^\uparrow|^2$ is absent in uniformly magnetized junctions without the RSO region \cite{zar1,majidi,chines}. 
In
Fig. \ref{DominantAAKR} bottom row, we change $\mu$ to
$0.5\Delta$ and $h=7.0\Delta$. Comparing panels ($a$)-($d$), we see that
by tuning the Fermi level one can have a great control over a
dominant reflection type at certain applied voltages. For instance, through our specific choice of
parameters' value, the anomalous Andreev is well separated from the
standard Andreev and spin-filliped normal reflections. Hence, at low voltage differences across the junction, one can access a
regime where the anomalous Andreev reflection is the highly dominated
reflection simply by tuning the Fermi level. This interesting regime can be determined through a conductance experiment that shall be discussed later. Also, the results can be understood by the band structure analysis presented above and Eq. (\ref{eqcang}). 
Another experimentally controllable parameter that a graphene layer
offers is the exertion of an external mechanical tension into the
graphene sheet. Our results have found that strain can also
change the Andreev and normal reflections the same as what was seen
for $\lambda$ and $\mu$. That is, the strain can render the system to a regime where the anomalous Andreev reflection dominates. We now proceed to discuss the generation of triplet pairs and the properties of charge and spin
conductances in a junction of F-RSO-S.    

     \begin{figure}[t]
     \includegraphics[width=8.cm, height=6.cm]{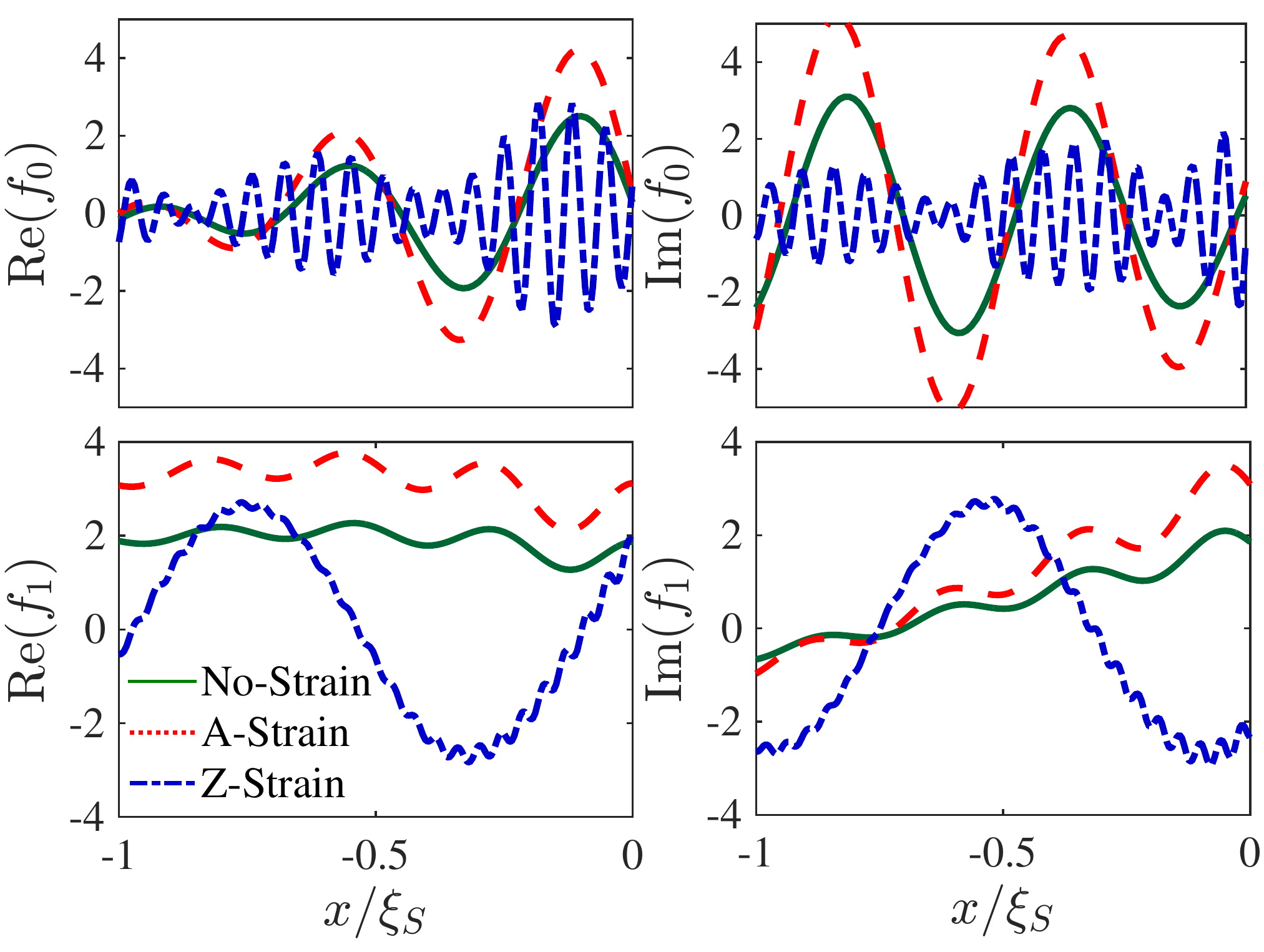}
     \caption{(Color online) Real and imaginary parts of OSP ($f_0$) and ESP ($f_1$) as a function of position in the F region $x \leq 0$ in the presence and absence of $A$- and $Z$-strains at a voltage difference of $eV=0.5\Delta$ across the contact. The chemical potential is also set at $\mu=5.0\Delta_0$. }
      \label{fig4}
      \end{figure}

\subsection{Equal- and opposite-spin pairings}\label{esp}
It is of fundamental
importance to find an experimentally feasible fashion to have control
over the amplitude and creation of the equal-spin triplets. Such a
control would help to unambiguously reveal the existence of such
odd-frequency superconducting correlations in a hybrid structure. As
discussed above, the RSO parameter $\lambda$ is experimentally
controlable through a simple external electric field \cite{ex1,ex2,ex3}. The
graphene layers have also provided a unique opportunity to tune the Fermi
level of a condensed matter system through applying a gate voltage. To further explore the relation between the anomalous Andreev reflections and triplet correlations, we calculate the opposite- and equal-spin pair correlations denoted by $f_0$ and $f_1$ \cite{halt1,halt2,halt3}. Following Ref. \onlinecite{halt1}, the OSP and ESP in the graphene system we consider can be expressed by:
\begin{SCfigure*}
\centering
\includegraphics[width=12.8cm,height=7.10cm]{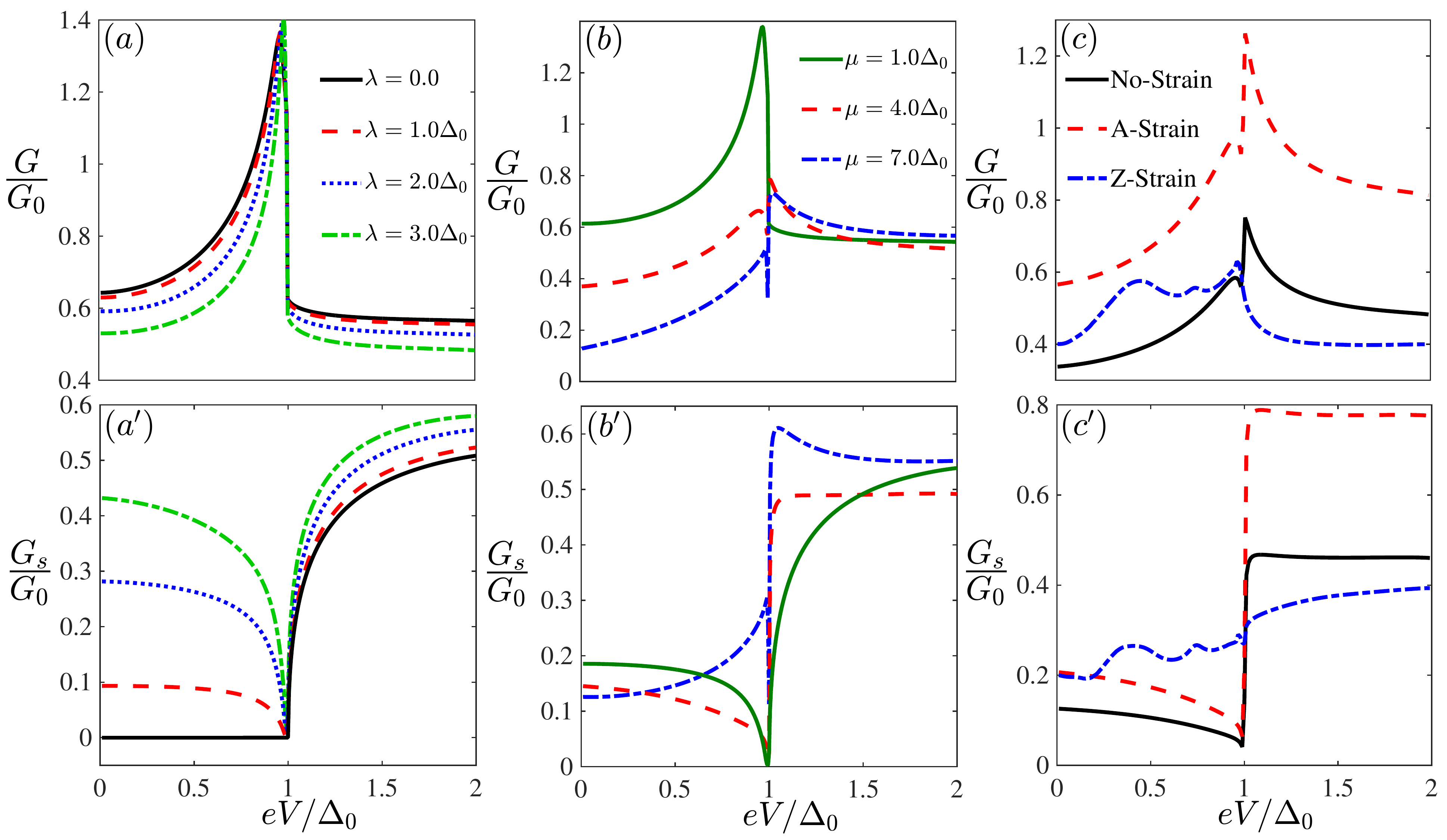}
    \caption{(Color online) ($a$)-($a'$) Normalized charge ($G$) and
      spin ($G_s$) conductances as a function of an applied voltage $eV$
      across the F-RSO-S junction for different values of the RSO
      parameter $\lambda=0.0, 1.0, 2.0$, and $3.0\Delta_0$. Here $G_0=G_{\uparrow}+G_{\downarrow}$ and we
      consider $\mu =\Delta_0, L=0.2\xi_S$, and
      $h=7.0\Delta_0$. ($b$)-($b'$) The normalized charge and spin
      conductances for various values of the chemical potential
      $\mu=1.0\Delta_0, 4.0\Delta_0$, and $7.0\Delta_0$ where
      $\lambda=1.5\Delta_0$ and $h=7.0\Delta_0$.  ($c$)-($c'$) The normalized charge and spin
      conductances for $A$-, $Z$-strain, and in the absence of strain at $\mu=5.0\Delta_0$,
      $h=7.0\Delta_0$, and $\lambda=1.5\Delta_0$.}
    \label{conduc}
\end{SCfigure*}
\begin{eqnarray}
&&\nonumber f_0(x,t)=\frac{1}{2}\sum_{\beta}\xi(t)[u^\uparrow_{\beta,K}v^{\downarrow*}_{\beta,K'}+u^\uparrow_{\beta,K'}v^{\downarrow*}_{\beta,K}-\\&& u^\downarrow_{\beta,K}v^{\uparrow*}_{\beta,K'}-u^\downarrow_{\beta,K'}v^{\uparrow*}_{\beta,K}],\\ &&\nonumber f_1(x,t)=-\frac{1}{2}\sum_{\beta}\xi(t)[u^\uparrow_{\beta,K}v^{\uparrow*}_{\beta,K'}+u^\downarrow_{\beta,K}v^{\downarrow*}_{\beta,K'}+\\&& u^\uparrow_{\beta,K'}v^{\uparrow*}_{\beta,K}+u^\downarrow_{\beta,K'}v^{\downarrow*}_{\beta,K}],
\end{eqnarray}
where $K$ and $K'$ denote different valleys and $\beta$ stands for A and B sublattices \cite{halt1}. Here
\begin{equation} 
\xi(t)=\cos(\varepsilon t)-i\sin(\varepsilon t)\tanh(\varepsilon/2T),
\end{equation} 
in which $t$ is the relative time in the Heisenberg picture \cite{halt1}. 
Figure \ref{fig3} illustrates the real and imaginary parts of OSP ($f_0$) and ESP  ($f_1$) near the F-RSO interface in the ferromagnetic region shown in Fig. \ref{fig1} i.e. $x\leq 0$. The different curves correspond to different Fermi levels equal to $\mu=1.0\Delta_0, 4.0\Delta_0, 7.0\Delta_0$. We set the voltage difference across the junction fixed at $eV=0.5\Delta_0$ whereas other parameters are identical to those of Fig. \ref{DominantAAKR}. When we set $\lambda=0$, the ESP $f_1$ vanishes and only $f_0$ remains nonzero. This is consistent with the known findings in a uniform ferromagnet coupled to a $s$ wave superconductor where the only nonvanishing triplet pairing is $f_0$ \cite{bergeret_rmp,Buzdin2005RMP,halt1,halt2,halt3}. A nonzero $\lambda$ however results in a finite nonvanishing ESP $f_1$ in addition to the presence of $f_0$. As seen in Fig. \ref{DominantAAKR}, for $\lambda\neq 0$, by calibrating $\mu$ one can highly suppress the amplitude of OSP $f_0$ while increase $f_1$. This finding is consistent with those of Ref. \onlinecite{halt1} for a F-S-F contact with noncolinear magnetization alignments. This is however in stark oppose to a conventional metal counterpart where the tunable chemical potential is absent \cite{halt1,equalspin1,equalspin2,halt3}. We note that the amplitude of anomalous Andreev reflection discussed in Fig. \ref{DominantAAKR} possesses identical aspects to the ESPs that demonstrates direct link of the equal-spin parings and the anomalous Andreev reflected holes.

We also introduce two kinds of strain: Strain applied along the Armchair direction ($A$-strain) and Zig-zag direction ($Z$-strain) i.e. $y$ and $x$ directions in Fig. \ref{fig1}, respectively. The applied tension into the graphene lattice causes anisotropic velocities in the $x$ and $y$ directions in each region $i$: $v_x^{i},v_y^{i}$ and aslo changes the coupling energy between carbon atoms $t_0 \sim 2.7$eV. To model the $A$- and $Z$-strain, we consider a strain strength of order of $\sim $ 20$\%$ which is equivalent to $s=0.2$ in our model (see Ref. \onlinecite{Alidoust2011}). In this amount of stress $s$, the coupling energies change to $t_1=t_2=0.96t_0$ and $t_3=0.5t_0$ for the $A$-strain while $t_1=t_2=0.56t_0$ and $t_3=1.1t_0$  for the $Z$-strain \cite{Alidoust2011}. Using these parameters, we calculate the corresponding anisotropic velocities of Dirac fermions in strained graphene lattice. Figure \ref{fig4} exhibits the real and imaginary parts of $f_0$ and $f_1$ pair correlations in $x\leq 0$ region for a strain-free junction and in the presence of $A$- and $Z$-strain. For simplicity in analysis, we set identical strains in each region, the chemical potential is fixed at $\mu=5.0\Delta_0$, exchange field $h=7.0\Delta_0$, $\lambda=1.5\Delta_0$, and a voltage difference at $eV=0.5\Delta_0$ across the junction. As seen, the amplitudes of OSPs $f_0$ and ESPs $f_1$ are highly influenced by the strain introduced. Interestingly, in the case of $Z$-strain, the amplitude of $f_0$ rapidly oscillates and diminishes while $f_1$ becomes less oscillating and smoother. This behaviour is suggestive of an experimentally feasible fashion to have control over the amplitude of both $f_0$ and $f_1$ so that one can suppress $f_0$ and simultaneously enhance $f_1$ in the same system. We have also investigated the back scattering amplitudes in the presence of strain (not shown). As one can expect, in the $Z$-strain mode, the anomalous Andreev reflection survives while the conventional Andreev reflection is vanishingly small that reaffirms the direct connection of $f_1$ and the anomalous equal-spin Andreev reflection.

\subsection{Charge and spin conductances}\label{cond}
An experimentally measurable quantity in such configuration is the junction conductance. 
Using the scattering coefficients, one can generalize the theory of Blonder-Tinkham-Klapwijk\cite{btk} to calculate the charge conductance via:
\begin{equation}
G=\int dq\sum_{\sigma,\sigma' =\uparrow,\downarrow}G_\sigma \Big(1- ‎|r_{N}^{\sigma'}|^2+|r_{A}^{\sigma}|^2 \Big),
\end{equation}
and the spin-polarized conductance:
\begin{equation}
G_s=\int dq \sum_{\sigma,\sigma'=\uparrow,\downarrow}G_\sigma \Big (1-|r_{N}^\sigma|^2+|r_{N}^{\sigma'}|^2‎
-|r_{A}^{\sigma'}|^2+|r_{A}^{\sigma}|^2 \Big),
\end{equation}
where $G_\sigma= 2e^2 |\varepsilon+\mu+\sigma h|$ and the junction
width is assumed enough wide so that one can replace
$\sum_n\rightarrow \int dq$. The behaviour of normalized spin and charge
conductances are shown in Fig. \ref{conduc}. We have normalized the conductances by $G_0=G_{\uparrow}+G_{\downarrow}$.
The presence of exchange energy $h$ in the F region results in an
imbalance between particles with different spin directions and causes
polarized currents at $eV \geq\Delta_0$ (black curve $\lambda=0$). In the presence of RSO
coupling $\lambda\neq 0$ however, due to the possibility of spin mixing in this
region, the spin-polarized conductance can be also nonzero in the subgap
region $eV \leq \Delta_0$. The exchange energy
and length of RSO region is kept fixed at $h=7.0\Delta_0$ and
$L=0.2\xi_S$, respectively. Panels ($a$) and ($a'$) illustrate the
role of RSO in the charge and spin conductances. As seen, by increasing
$\lambda$, the spin subgap conductance drastically enhances
particularly at zero voltage bias while the associated
charge conductance decreases. The enhancement of spin subgap
conductance traces back to the generation of equal-spin triplet
correlations (see Fig. \ref{fig3}, corresponding triplet correlations). Figure \ref{conduc} also shows
the effect of chemical potential in panels ($b$) and ($b'$). While the
chemical potential can reduce the charge subgap conductance, the spin subgap
conductance remains about the same at low voltages for the chosen set of parameters. The changes in the spin-polarized conductance is more pronounced at larger voltages. The influence of $A$- and
$Z$-strain is shown in ($c$) and ($c'$). We see that in the $Z$- strain mode the spin subgap conductance is almost largest while the corresponding charge conductance is less than $A$-strain mode. The conductance behaviours are consistent
with the associated backscattering probabilities and superconducting triplet correlations presented in the previous subsections.

\section{conclusion}\label{concld}

In conclusion, motivated by recent experiments \cite{avsar2014nat,ex1,ex2,ex3}, we have theoretically studied backscattering probabilities, triplet superconducting correlations, and charge/spin conductance in a ferromagnet-Rashba spin orbit-superconductor graphene-based junction. Our findings offer an experimentally feasible platform for creating tunable superconducting equal-spin triplet correlations. The odd-frequency triplet correlations are formed near the F-RSO interface through an anomalous equal-spin Andreev backscattering. Considering the band structure of system, the anomalous reflection is allowed due to a band splitting in the presence of Rashba spin orbit coupling in the RSO region. We show that the amplitude of equal-spin pair correlations can be enhanced by varying the Fermi level in nonsuperconducting region, exertion of strain into the graphene layer, and controlling the strength of RSO through an external electric field while the amplitude of opposite-spin pair correlations suppressed simultaneously. The anomalous equal-spin Andreev reflection also causes a nonzero spin-polarized \textit{subgap} conductance.
These phenomena can be revealed in a conductance spectroscopy experiment. More importantly, the signatures of the equal-spin pairings on the experimentally observable quantities discussed here can supply suitable insights into the proximity-induced Rashba spin orbit couplings recently achieved in experiments \onlinecite{avsar2014nat,ex1,ex2,ex3}. 

\acknowledgements 
We thank M. Salehi for numerous helpful conversations. M.A. also would like to thank K. Halterman for useful discussions.

\begin{widetext}
\section{appendix}

The wavefunctions associated with the dispersion relation in the F region are:
  \begin{equation}
 \begin{array}{l}
\psi^{\text{F},\pm}_{e,\uparrow}(x)=\left(\mathbf{0^2}, 1 , \pm e^{\pm i
    \alpha_\uparrow^e}, \mathbf{0^4}\right)^{\bf T} e^{\pm i
  k_{x,\uparrow}^{\text{F},e} x},\\
\\
\psi^{\text{F},\pm}_{e,\downarrow}(x)=\left( 1 , \pm e^{\pm i \alpha_\downarrow^e}, \mathbf{0^2},\mathbf{0^4}\right)^{\bf T} e^{\pm i k_{x,\downarrow}^{\text{F},e} x},
\\
\\
   \psi^{\text{F},\pm}_{h,\uparrow}(x)= \left(\mathbf{0^4}, 1 ,  \mp
     e^{ \pm i \alpha_\uparrow^h},\mathbf{0^2}\right)^{\bf T} e^{ \pm
     i k_{x,\uparrow}^{\text{F},h} x},\\
\\
 \psi^{\text{F},\pm}_{h,\downarrow}(x)=\left( \mathbf{0^4},
   \mathbf{0^2}, 1 , \mp e^{ \pm i\alpha_\downarrow^h}\right)^{\bf T}
 e^{\pm i k_{x,\downarrow}^{\text{F},h} x},
  \end{array}
     \end{equation} 
where $\mathbf{0^n}$ represents a $1 \times n$ matrix with only zero
entries and ${\bf T}$ is a transpose operator. We assume that the
junction width $W$ is enough large so that the $y$ component of
the wavevector $k_y$ is a conserved quantity upon the scattering
processes and therefore, we factored out the corresponding
multiplication i.e. $\exp(i k_y y)$. 
The $\alpha_{e,(h)}^{\uparrow(\downarrow)}$ variables are the
propagation angles in the presence of strain and are given by:
   \begin{equation}
   \alpha_{\uparrow, (\downarrow)}^{e(h)}=\arctan\left(\frac{v_y^\text{F} q_n}{v_x^\text{F} k_{x, \uparrow(\downarrow)}^{\text{F},e(h)}} \right).
   \end{equation}
The $e(h)$ superscript indicates electron (hole)-like parameters
and $\uparrow (\downarrow)$ subscript denotes the spin orientation. 
The $x$ component of wavevectors are not conserved during the
scattering processes and can be expressed by:
   \begin{equation}
    \begin{cases}
   k_{x,\uparrow}^{\text{F},e}=(\hbar v_x^\text{F})^{-1}\left(\varepsilon+\mu^\text{F}+h\right)\cos\alpha_\uparrow^e\\
\\
    k_{x,\downarrow}^{\text{F},e}=(\hbar v_x^\text{F})^{-1}\left(\varepsilon+\mu^\text{F}-h\right)\cos\alpha_\downarrow^e\\
\\
    k_{x,\uparrow}^{\text{F},h}=(\hbar v_x^\text{F})^{-1}\left(\varepsilon-\mu^\text{F}-h\right)\cos\alpha_\uparrow^h\\
\\
    k_{x,\downarrow}^{\text{F},h}=(\hbar v_x^\text{F})^{-1}\left(\varepsilon-\mu^\text{F}+h\right)\cos\alpha_\downarrow^h
    \end{cases}.
    \label{Eq.10}
    \end{equation} 
We denote $k_y^i\equiv q_n$ that can vary in interval $-\infty \leq
q_n \leq +\infty$. The $x$ component of the wavevector however becomes
imaginary for larger values of $q_n$ than a critical value
$q^c$. The wavefunctions for $q_n>q^c$ are decaying functions and
therefore, depending on the junction geometry, are not able to contribute to the transport process. 
The critical values can be expressed as follows:   
   \begin{equation}
   \begin{cases}
   q_{e,\uparrow}^c=(\hbar v_y^\text{F})^{-1}\left|\varepsilon+\mu^\text{F}+h\right| \\
   \\
   q_{e,\downarrow}^c=(\hbar v_y^\text{F})^{-1}\left|\varepsilon+\mu^\text{F}-h\right| \\
   \\
   q_{h,\uparrow}^c=(\hbar v_y^\text{F})^{-1}\left|\varepsilon-\mu^\text{F}-h\right|\\
   \\
   q_{h,\downarrow}^c=(\hbar v_y^\text{F})^{-1}\left|\varepsilon-\mu^\text{F}+h\right|
   \end{cases}.
   \label{Eq.11}
   \end{equation}

In contrast to the intrinsic spin orbit couplings, the energy spectrum in
the presence of RSO is gapless with a splitting of magnitude
$2\lambda$ between subbands in the RSO region. 
The wavefunctions associated with the eigenvalues given in
the text can be expressed by: 
  \begin{equation}
  \begin{array}{l}
  \psi^{\text{RSO},\pm}_{e,\eta=+1}(x)= \Big (\mp i f_{+}^e e^{\mp i
      \theta_{+}^e}, -i, 1, \pm f_{+}^e e^{\pm i
      \theta_{+}^e},\mathbf{0^4}\Big)^{\bf T}e^{\pm i k_{x,+}^{\text{RSO},e}x}\\
  \\
  \psi^{\text{RSO},\pm}_{e, \eta=-1}(x)=\Big(\pm f_{-}^e e^{\mp i \theta_{-}^e}, 1, -i, \mp if_{-}^e e^{\pm i \theta_{-}^e},\mathbf{0^4}\Big)^{\bf T}e^{\pm i k_{x,-}^{\text{RSO},e}x}\\
  \\
  \psi^{\text{RSO},\pm}_{h,\eta=+1}(x)= \Big (\mathbf{0^4},\mp i f_{+}^h e^{\mp i \theta_{+}^h}, -i, 1, \pm f_{+}^h e^{\pm i \theta_{+}^h}\Big)^{\bf T}e^{\pm i k_{x,+}^{\text{RSO},h}x}\\
  \\
  \psi^{\text{RSO}, \pm}_{h, \eta=-1}(x)=\Big (\mathbf{0^4}, \pm f_{-}^h e^{\mp i \theta_{-}^h}, 1, -i, \mp if_{-}^h e^{\pm i \theta_{-}^h}\Big)^{\bf T}e^{\pm i k_{x,-}^{\text{RSO},h}x}\\
  \end{array}.
  \label{Eq.13}
  \end{equation}
Here also the transverse component of wavevector is factored out due
to the conservation discussion made earlier. The $x$ component of
the wavevector however is not conserved during the scattering
processes:   
 \begin{equation}
   \begin{cases}
   k_{x,\eta}^{\text{RSO},e}=(v_x^\text{RSO})^{-1}(\mu^\text{RSO}+\varepsilon)f_{\eta}^e\cos\theta_\eta^e\\
\\
   k_{x,\eta}^{\text{RSO},h}=(v_x^\text{RSO})^{-1} (\mu^\text{RSO}-\varepsilon)f_{\eta}^e\cos\theta_\eta^h
   \end{cases},
   \label{Eq.14}
   \end{equation}
and the definition of auxiliary parameters are:   
  \begin{equation}
  \begin{cases}
  f_\eta^e=\sqrt{1+2\eta\lambda(\mu^\text{RSO}+\varepsilon)^{-1}} \\
\\
  f_\eta^h=\sqrt{1+2\eta\lambda(\mu^\text{RSO}-\varepsilon)^{-1}}\\
  \end{cases}
  \label{Eq.15},\;\;\;
\begin{cases}
      \theta_{\eta}^e=\arctan\left(\frac{q_n  v_y^\text{RSO}
        }{v_x^\text{RSO} k_{x, \eta}^{\text{RSO},e}}\right) \\
\\
       \theta_{\eta}^h=\arctan\left(\frac{q_n  v_y^\text{RSO}}{v_x^\text{RSO} k_{x, \eta}^{\text{RSO},h}}\right)\\
  \end{cases},
  \end{equation} 
where $\theta_{\eta}^{e(h)}$ are the electron and hole propagation
angles in the region with spin orbit interaction.
We note that if the transverse component of wavevector goes beyond a
critical value $q^c$, the same as what discussed for the ferromagnet region,
the wavefunctions turn to evanescent modes. Here, however, since the
RSO region is sandwiched between F and S regions, the evanescent modes
contribute to the quantum transport process. We thus take both the
propagating and decaying modes into account throughout our calculations.  
  
In the superconductor region, $U_0$ denotes the electrostatic
potential that is very large ($U_0\gg 1$) in actual experiments
compared to other system energy scales so that the
step function assumption made above for the pair potential can be a good
approximation in numerous realistic cases. 
The wavefunctions in the superconducting region are given by:
   \begin{equation}
   \begin{array}{l}
   \psi^{\text{S},\pm}_{e, 1}(x)=\Big(e^{+i \beta}, \pm e^{+i
     \beta}, \mathbf{0^2}, e^{-i\phi}, \pm e^{-i\phi},
   \mathbf{0^2}\Big)^{\bf T}e^{\pm i k_x^{\text{S},e} x}\\
   \\
   \psi^{\text{S},\pm}_{e,2}(x)=\Big (\mathbf{0^2}, e^{+i \beta}, \pm e^{+i \beta}, \mathbf{0^2}, e^{-i\phi}, \pm e^{-i\phi}\Big)^{\bf T}e^{\pm i k_x^{\text{S},e} x}\\
   \\
   \psi^{\text{S},\pm}_{h, 1}(x)= \Big (e^{-i \beta}, \mp e^{-i \beta}, \mathbf{0^2}, e^{-i\phi}, \mp e^{-i\phi}, \mathbf{0^2}\Big)^{\bf T}e^{\pm i k_x^{\text{S},h} x}\\
   \\
   \psi^{\text{S}, \pm}_{h, 2}(x)= \Big (\mathbf{0^2}, e^{-i \beta}, \mp e^{-i \beta}, \mathbf{0^2}, e^{-i\phi}, \mp e^{-i\phi}\Big)^{\bf T}e^{\pm i k_x^{\text{S},h} x}
   \end{array}
   \label{WF.S}.
   \end{equation} 
The parameter $\beta$ is responsible for the electron-hole conversions at
the interface RSO-S and depends on the superconducting gap: 
    \begin{equation}
    \beta=\begin{cases}
    +~\arccos(\varepsilon/\Delta_0)&       \varepsilon\leq \Delta_0 \\
    -i ~\text{arccosh}(\varepsilon/\Delta_0) &   \varepsilon \geq\Delta_0
    \end{cases}
    \label{Eq.19}.
    \end{equation}
Similar expressions to those found for the longitudinal component of
wavevector in the F and RSO region appear for the S. 
In the F region, we assume that a right moving electron with spin-up
direction hits the interface of F-RSO with energy $\varepsilon$. This
particle can reflect back as: (a) an electron with spin-up direction
(conventional normal reflection), (b) as a hole with spin-down
direction (conventional Andreev reflection), (c) as an electron with
spin-down direction (spin flipped normal reflection), and (d) as a
hole with spin-up direction (anomalous Andreev reflection). Thus, the total wavefunction in the F region can be written as:
  \begin{equation}
  \begin{array}{rl}
  \Psi^\text{F}(x)=& \left(\mathbf{0^2}, 1 , e^{+i \alpha_\uparrow^e},
    \mathbf{0^4}\right)^{\bf T} e^{+i k_{x,\uparrow}^{\text{F},e} x} +
  r_{N}^\uparrow \left(\mathbf{0^2}, 1 , -e^{-i \alpha_\uparrow^e},
    \mathbf{0^4}\right) ^{\bf T} e^{-i k_{x,\uparrow}^{\text{F},e} x}+
  r_{N}^\downarrow \left( 1 , -e^{-i
      \alpha_\downarrow^e},\mathbf{0^2}, \mathbf{0^4}\right) ^{\bf T} e^{-i
    k_{x,\downarrow}^{\text{F},e} x}+\\
\\
  &+ r_{A}^\uparrow \left(\mathbf{0^4}, 1 , e^{-i \alpha_\uparrow^h},\mathbf{0^2}\right) ^{\bf T} e^{-i k_{x,\uparrow}^{\text{F},h} x}+ r_{A}^\downarrow \left( \mathbf{0^4}, \mathbf{0^2}, 1 , e^{-i \alpha_\downarrow^h}\right) ^{\bf T} e^{-i k_{x,\downarrow}^{\text{F},h} x}.
  \end{array}
  \label{WF.F}
  \end{equation}
where $r_{N}^\uparrow$, $r_{N}^\downarrow$, $r_{A}^\downarrow$, and
$r_{A}^\uparrow$ are the amplitudes of conventional, anomalous normal
reflections, conventional, and anomalous Andreev reflections,
respectively. 
When an electron hits the F-RSO interface, it can enter
into the RSO region through one of its subbands and reflect back as an
electron or hole upon
collision with the RSO-S interface. Each subband is a mixture of spin-up and -down due to the presence of RSO coupling.
Hence, in the RSO region, $0 \leq x \leq L$, the total wavefunction is
given by:
\begin{equation}
\begin{array}{rl}
\Psi^{\text{RSO}}(x) &= a_1 \Big(-i f_{+}^e e^{-i \theta_{+}^e}, -i, 1, f_{+}^e e^{i \theta_{+}^e},\mathbf{0^4}\Big) ^{\bf T} e^{i k_{x,+}^{\text{RSO},e}x} + a_2 \Big(i f_{+}^e e^{i \theta_{+}^e}, -i, 1, -f_{+}^e e^{-i \theta_{+}^e},\mathbf{0^4}\Big)^{\bf T} e^{-i k_{x,+}^{\text{RSO},e}x}\\
&+ a_3 \Big(f_{-}^e e^{-i \theta_{-}^e}, 1, -i, -if_{-}^e e^{i \theta_{-}^e},\mathbf{0^4}\Big)^{\bf T} e^{i k_{x,-}^{\text{RSO},e}x}+ a_4 \Big(- f_{-}^e e^{i \theta_{-}^e}, 1, -i, if_{-}^e e^{-i \theta_{-}^e},\mathbf{0^4}\Big)^{\bf T} e^{-i k_{x,-}^{\text{RSO},e}x}\\
&+ a_5 \Big(\mathbf{0^4},-i f_{+}^h e^{-i \theta_{+}^h}, -i, 1, f_{+}^h e^{i \theta_{+}^h}\Big) ^{\bf T} e^{i k_{x,+}^{\text{RSO},h}x}+ a_6 \Big(\mathbf{0^4},i f_{+}^h e^{i \theta_{+}^h}, -i, 1, -f_{+}^h e^{-i \theta_{+}^h}\Big) ^{\bf T} e^{-i k_{x,+}^{\text{RSO},h}x}\\
&+ a_7 \Big(\mathbf{0^4}, f_{-}^h e^{-i \theta_{-}^h}, 1, -i, -if_{-}^h e^{i \theta_{-}^h}\Big) ^{\bf T} e^{i k_{x,-}^{\text{RSO},h}x}+ a_8 \Big(\mathbf{0^4}, -f_{-}^h e^{i \theta_{-}^h}, 1, -i, if_{-}^h e^{-i \theta_{-}^h}\Big) ^{\bf T} e^{-i k_{x,-}^{\text{RSO},h}x}.\\
\end{array}
\label{WF.RSO}
\end{equation}
As seen, the total wavefunction in this region involves 8 unknown
coefficients $a_{1,\dots ,8}$ for spin-up and -down particles and holes. 
Finally, the total wavefunction in the S region can be written as follows:
 \begin{equation}
 \begin{array}{rl}
 \Psi^\text{S}(x)&= t_1 \Big(e^{i \beta}, e^{i \beta}, \mathbf{0^2}, e^{-i\phi}, e^{-i\phi}, \mathbf{0^2}\Big) ^{\bf T} e^{i k_x^\text{S} x} +t_2 \Big(\mathbf{0^2}, e^{i \beta}, e^{i \beta}, \mathbf{0^2}, e^{-i\phi}, e^{-i\phi}\Big) ^{\bf T} e^{i k_x^\text{S} x}\\
 &+ t_3 \Big(e^{-i \beta}, -e^{-i \beta}, \mathbf{0^2}, e^{-i\phi}, -e^{-i\phi}, \mathbf{0^2}\Big) ^{\bf T} e^{-i k_x^\text{S} x} +t_4 \Big(\mathbf{0^2}, e^{-i \beta}, -e^{-i \beta}, \mathbf{0^2}, e^{-i\phi}, -e^{-i\phi}\Big) ^{\bf T} e^{-i k_x^\text{S} x}
 .\end{array}
 \label{WF.S}
 \end{equation} 
Here, the transmission coefficients are denoted by $t_{1,2,3,4}$. The
macroscopic phase of superconductivity $\phi$ plays no role in the
geometry considered and thus we set it zero. In the above wave
function, we assumed that the S region is in a heavily dopped regime i.e. $U_0 \gg \varepsilon,\Delta_0 $.
By matching
 the wavefunctions at the interfaces, i.e.,
 $\Psi^\text{F}(x)=\Psi^\text{RSO}(x)$ at $x=0$ and $\Psi^\text{RSO} (x)=\Psi^\text{S} (x)$ at $x=L$, we
 obtain the unknown scattering coefficients. The resulting
 coefficients however are very large and complicated expressions and we skip to
 present them. 
 
\end{widetext}

\end{document}